\documentstyle[preprint,aps,epsfig,tighten]{revtex}

\begin{document}
\title{THE $p(d,d^{\prime })$ REACTION AND THE $\sigma NN^{\ast }(1440)$ COUPLING
CONSTANT}
\author{B. Juli\'{a}-D\'{\i}az$^{(1)}$, A. Valcarce$^{(2)}$, F. Fern\'{a}ndez$^{(2)}$}
\address{$^{(1)}$ Department of Physical Sciences, University of Helsinki and\\
Helsinki Institute of Physics,\\
P.O. Box 64, 00014 Helsinki, Finland}
\address{$^{(2)}$ Grupo de F\' \i sica Nuclear,\\
Universidad de Salamanca, E-37008 Salamanca, Spain}
\maketitle

\begin{abstract}
We make use of a $NN \to NN^*(1440)$ transition potential derived 
from a quark-model in a parameter-free way, 
to study the Roper excitation diagram contributing 
to the reaction $p(d,d^{\prime})$. We also determine
the $\pi NN^*(1440)$ and $\sigma NN^*(1440)$ coupling
constants.
\end{abstract}

\section{Introduction}

The $N^{\ast }(1440)$ (Roper) is a broad resonance which couples strongly (60%
$-$70$\%$) to the $\pi N$ channel and significantly (5$-$10$\%$) to the $%
\sigma N$ channel \cite{1}. It may therefore play an
important role in nuclear dynamics as an intermediate state. Graphs
involving the excitation of the $N^{\ast }(1440)$ appear in many different
reactions from the $p(\alpha ,\alpha ^{\prime })$ to the heavy ion
collisions at relativistic energies \cite{2,3,4,5,6,7}

Usually the $NN\rightarrow NN^{\ast }(1440)$ transition potential is taken
as a straightforward generalization of some pieces of the $NN$ interaction,
scaling the coupling constants and incorporating resonance
width effects. However this procedure may have serious shortcomings
specially concerning the short-range part of the interaction \cite{8}.

In this talk we present some applications of a recently derived $%
NN\rightarrow NN^{\ast }(1440)$ transition potential\cite{9}, obtained by
means of the same quark-model approach previously used to study the $NN$
system and transition potentials involving the $\Delta$ 
resonance \cite{8,10}. A main
feature of the quark treatment is its universality in the sense that all the
baryon-baryon interactions are treated on an equal footing. Therefore, once
the model parameters are fixed from $NN$ data there are no free parameters
for any other case. A second important aspect is the appearance of quark
exchanges between baryons, coming from quark antisymmetry. As quarks cannot
be exchanged between two baryons if their wave functions do not overlap, the
exchange contributions are necessarily short ranged.

After a brief description of the quark model based $NN\rightarrow NN^{\ast
}(1440)$ interaction (a complete discussion can be found in \cite{9}), we
center our attention in the study of a reaction mediated by the excitation
of the Roper resonance, the $p(d,d^{\prime })$ reaction, and on the
derivation of the $\pi NN^{\ast }(1440)$ and $\sigma NN^{\ast }(1440)$
coupling constants.

\section{$NN\rightarrow NN^{\ast }(1440)$ transition potential}

In the Born-Oppenheimer approximation the $NN\rightarrow NN^{\ast }(1440)$
potential at interbaryon distance $R$ is obtained by sandwiching the $qq$
potential between $NN$ and $NN^{\ast }(1440)$ states, written in terms of
quarks, for all the pairs formed by two quarks belonging to different
baryons. In the model of Ref. \cite{9} the $qq$ potential contains a
confining term taken to be linear ($r_{ij}$), the usual perturbative
one-gluon-exchange (OGE) interaction containing Coulomb ($1/r_{ij}$),
spin-spin (${\vec{\sigma}}_{i}\cdot {\vec{\sigma}}_{j})$ and tensor ($S_{ij}$%
) terms, and pion (OPE) and sigma (OSE) exchanges 
as a consequence of the breaking of
chiral symmetry. The $N^{\ast }(1440)$ and $N$ are given by $|N^{\ast
}(1440)\rangle =\left\{ \sqrt{\frac{2}{3}}|[3](0s)^{2}(1s)\rangle -\sqrt{%
\frac{1}{3}}|[3](0s)(0p)^{2}\rangle \right\} \otimes \lbrack 1^{3}]_{c}$ and 
$|N\rangle =|[3](0s)^{3}\rangle \otimes \lbrack 1^{3}]_{c}$, where $%
[1^{3}]_{c}$ is the completely antisymmetric color state, $[3]$ is the
completely symmetric spin-isospin state and $0s$, $1s$, and $0p$, stand for
harmonic oscillator orbitals.

\begin{figure}[h!]
\begin{center}
\mbox{\epsfxsize=70mm\epsffile{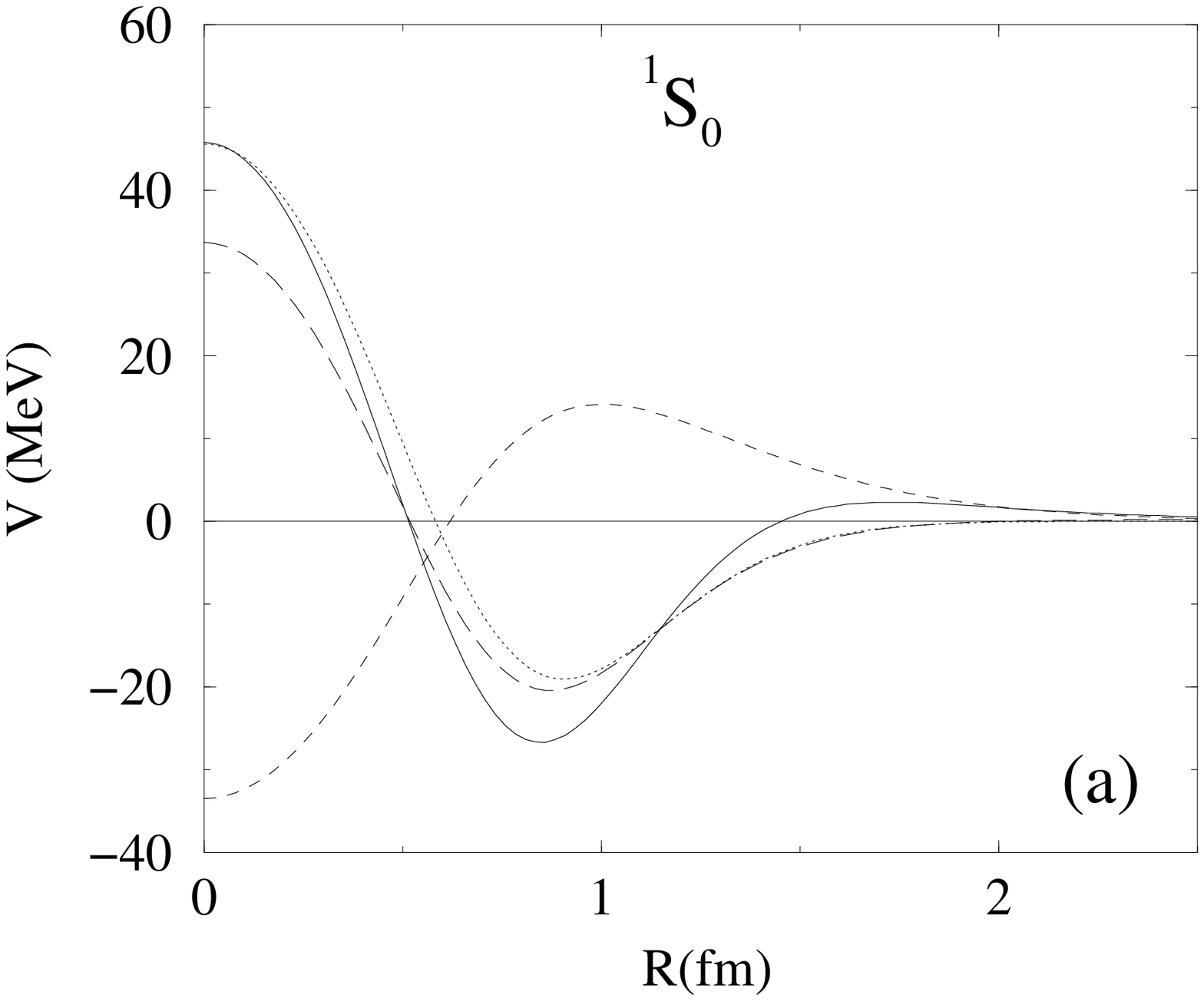}}
\mbox{\epsfxsize=70mm\epsffile{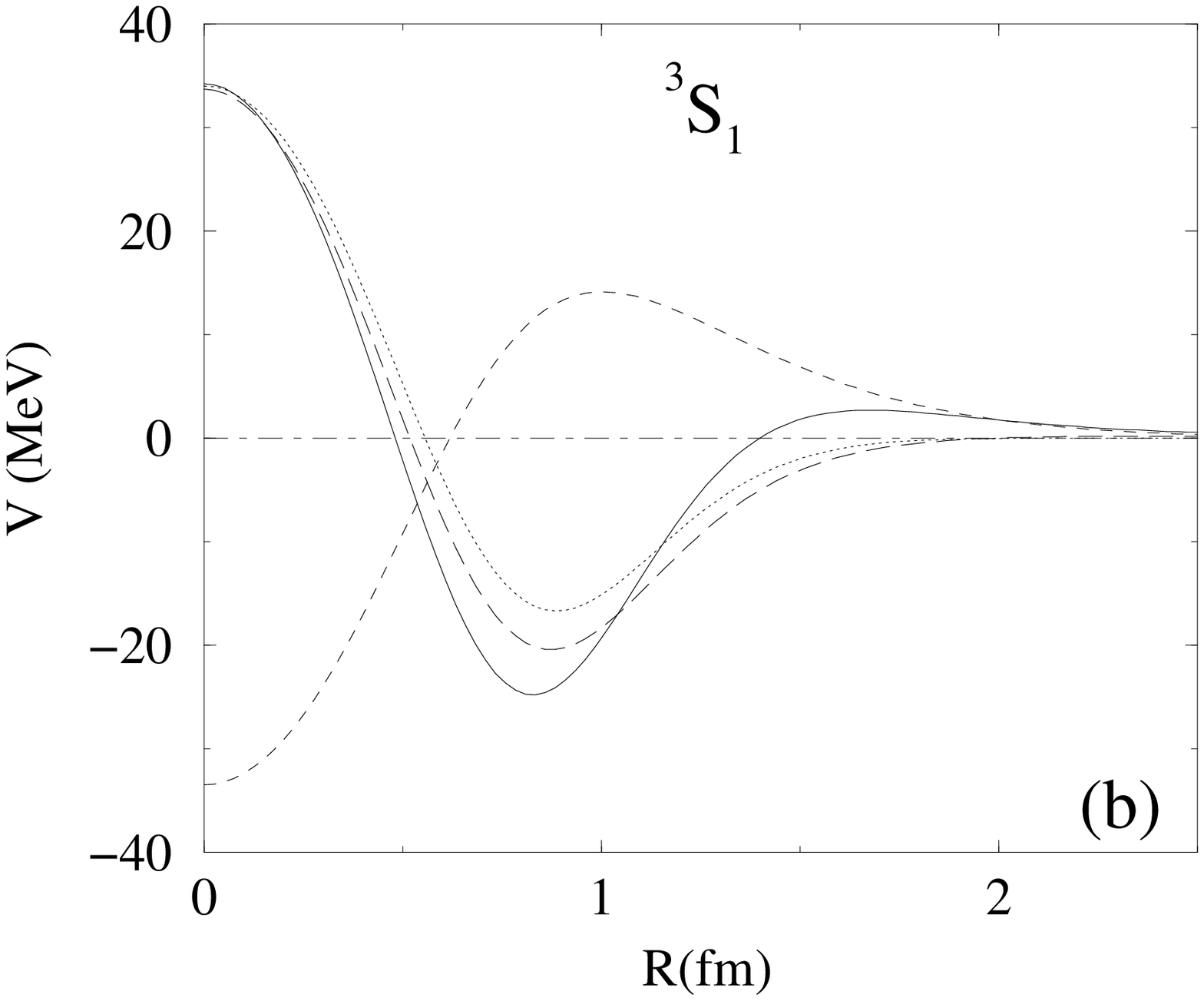}}
\end{center}
\caption{$NN \to NN^*(1440)$ potential 
for (a) the $^1S_0$ partial wave, and (b) the $^3S_1$
partial wave. We have denoted by
the long-dashed, dashed, dotted, and dot-dashed lines, 
the central OPE, OSE, OGE, and the
tensor contributions, respectively. 
By the solid line we plot the total potential.}
\label{fig1}
\end{figure}

In Fig. \ref{fig1} we show the potentials 
obtained for $L=0$ $(^{1}S_{0}$ and 
$^{3}S_{1})$. The interaction is repulsive 
at short range $(0<R \,\, ({\rm fm})<0.6)$,
attractive at intermediate range $(0.6<R \,\, ({\rm fm}) <1.4)$, 
and asymptotically repulsive,
this last behavior contrary to the naive expectation. This sign reversal is
a direct consequence of the presence of a node in the\ $N^{\ast }(1440)$
wave function which implies a change of sign with respect to the $N$ wave
function. Certainly it is possible to choose the opposite sign for the $%
N^{\ast }(1440)$ wave function with respect the $N$, and in this case the
long-range part of the transition potential would become attractive but
there would be also an unexpected change from repulsion to attraction at
short-range. This makes evident that the $NN\rightarrow NN^{\ast }(1440)$
transition potential looks very different from the $NN$ interaction and that
the simple scaling procedure seems not to be appropriate to derive the
interaction with a resonance.

\section{Roper excitation in \lowercase{$pd$} scattering}

There are two experiments where the $N^{\ast }(1440)$ resonance contribution
has been isolated by means of model-dependent theoretical methods.
The first one is the $p(\alpha ,\alpha ^{\prime })$ reaction carried out in
Saclay \cite{11} already ten years ago. The data showed two peaks in the
cross section, the most prominent one attributed to a $\Delta $ excitation in
the projectile (DEP) \cite{12}, and the second one explained as a Roper
excitation in the target (RET) \cite{3}. The second experiment is the $%
p(d,d^{\prime })$ reaction, that was studied making use of the same
mechanisms \cite{13}. These two reactions are particularly interesting
because in both cases the projectile ($d$ or $\alpha $) has $T=0$. This
ensures that the $N^{\ast }(1440)$ reaction mechanism can only be driven by
a scalar interaction.

Our purpose in this section is the study the target Roper excitation process
in the $p(d,d^{\prime })$ reaction making use of the quark model $%
NN\rightarrow NN^{\ast }(1440)$ transition potential.
We will consider the data where the $\Delta$
contribution has been subtracted \cite{13} as our experimental data.
The amplitude for the elementary process of $N^{\ast }(1440)$ production can
be written in terms of the scalar transition potential $(V_{0})_{NN%
\rightarrow NN^{\ast }}$ as \cite{13}: 
\begin{equation}
|M|^{2}=12F_{d}^{2}\left( {\frac{f^{\prime }}{m_{\pi }}}\right) ^{2}|G^{\ast
}|^{2}\left| (V_{0})_{NN\rightarrow NN^{\ast }}(q_{cm})\right|
^{2}q_{cm}^{2}\,.
\end{equation}
The function $F_{d}(\vec{k})$ is the deuteron form factor 
\begin{equation}
F_{d}(\vec{k})=\int d\vec{r}\;\phi ^{\ast }(\vec{r})\;e^{i{\frac{\vec{k}%
\cdot \vec{r}}{2}}}\;\phi (\vec{r})
\end{equation}
where $\phi (\vec{r})$ is the deuteron S-wave function, and the momentum $%
\vec{k}=\vec{p}_{d}-\vec{p}_{d^{\prime }}$ is taken in the initial deuteron
rest frame. $q_{cm}$ is the momentum transfer between the nucleons in the
center of mass system and $f^{\prime }\equiv f_{\pi NN^{\ast }}$. $G^{\ast }$
is the $N^{\ast }(1440)$ propagator as given in Ref. \cite{13}.

\begin{figure}[h!]
\begin{center}
\vspace{15pt} \mbox{\epsfxsize=77mm \epsfysize=67mm\epsffile{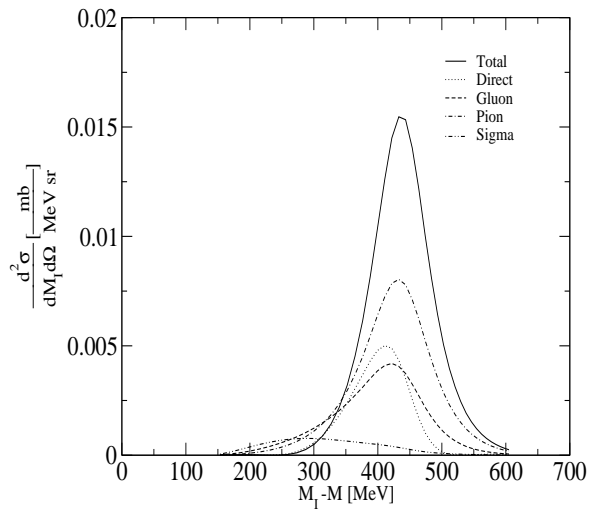}}
\end{center}
\par
\centering
\caption{Detailed contributions to the $p(d,d')$ cross section coming from the
different interactions at the quark level, neglecting the interference
terms. We denote by direct the result obtained neglecting quark-exchange
diagrams. $M_I$ is the invariant mass of the target system.}
\label{fighire1}
\end{figure}

\begin{figure}[h!]
\begin{center}
\vspace{18pt} \mbox{\epsfxsize=80mm \epsfysize=70mm\epsffile{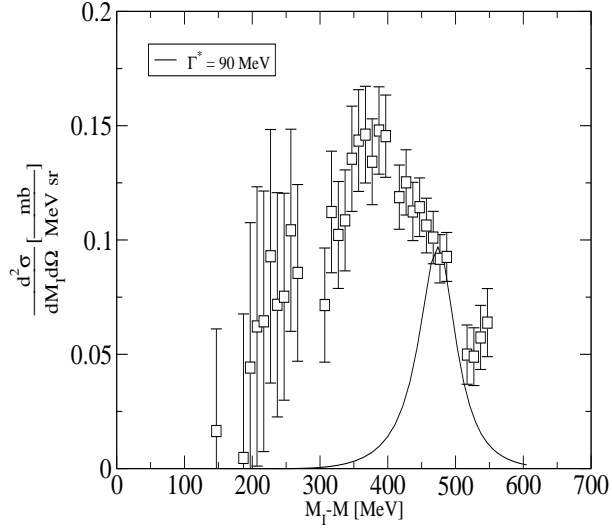}}
\end{center}
\par
\centering
\caption{Quark model result for the RET process contributing to the
$p(d,d')$ reaction.
$M_I$ is the invariant mass of the target system.
Experimental data correspond to $T_d=$ 2.3 GeV
and $\theta^L=$ 1.1 deg. They were obtained in Ref. \protect\cite{13}
by means of a theoretical subtraction of the $\Delta$ contribution.}
\label{fighire}
\end{figure}

In order to perform the calculation, we need to extract the genuine scalar
potential at all distances from the quark-model based $NN\rightarrow
NN^{\ast }(1440)$ interaction. Such potential presents a
non-trivial structure at short distances due to the quark antisymmetrizer,
which involves operators of the type $(1+\vec{\sigma}_{i}\vec{\sigma}_{j})$
and $(1+\vec{\tau}_{i}\vec{\tau}_{j})$. When combined with the corresponding
spin-isospin operators of each piece of the interaction one obtains a
general form $V_{NN\rightarrow NN^{\ast }}^{(S,T)}=V_{0}+V_{1}(\;\vec{\sigma
_{1}}\cdot \vec{\sigma _{2}})+V_{2}\;(\vec{\tau _{1}}\cdot \vec{\tau _{2}}%
)+V_{3}\;(\vec{\sigma _{1}}\cdot \vec{\sigma _{2}})\;(\vec{\tau _{1}}\cdot 
\vec{\tau _{2}})$, after projecting the quark spin-isospin degrees of freedom
into nucleonic spin-isospin degrees of freedom \cite{14}. From this projection,
the functions $V_{i}$ can be easily calculated \cite{15}
In Fig. \ref{fighire1} we present the 
contribution to the $p(d,d^{\prime })$ cross
section coming from the different interactions at quark level. The most
important ones are those from the scalar pieces generated from the pion and
gluon exchange combined with quark exchanges. This shows that the process is
driven by the short-range part of the interaction, where quark exchanges are
relevant. In Fig. \ref{fighire} we 
compare the result obtained using the quark-model
derived $NN\rightarrow NN^{\ast }(1440)$ potential to the experimental data.
As can be seen, the cross section is understimated, coming closer to data if
one chooses a smaller value for the $N^{\ast }(1440)$ width. The bigger
disagreement with the extracted data corresponds to the region where the
error bars are larger, in other words, to the region where the uncertainties
related to the theoretical method used to subtract the $\Delta $
contribution and interference term are important. The subtraction of the 
$\Delta $ contribution is proportional to the square of the $\pi N\Delta $
coupling constant. Its value is different in baryonic processes, $f_{\pi
N\Delta }^{2}/4\pi =0.35$, than the one used in our model, $f_{\pi N\Delta
}^{2}/4\pi =0.22$ \cite{16}, because it includes tensor coupling
between the $^{1}S_{0}$ $NN$ and the $^{5}D_{0}$ $N\Delta $ partial waves. As
a consequence, the baryonic calculation of the $\Delta $ contribution could
be underestimating the region above the peak overestimating in this way the 
$N^{\ast }(1440)$ contribution. The way to wipe out those uncertainties would
be to calculate the $\Delta $ contribution together with the interference
term making use of quark-model potentials.

\section{$\protect\pi NN^{\ast }(1440)$ and $\protect\sigma NN^{\ast }(1440)$
coupling constants.}

The usual way to determine meson$-NN$ coupling constants is through the
fitting of $NN$ scattering data with phenomenological meson exchange models.
However $NN$ $^{\ast }$ scattering data do not exist and one has to resort
to another procedure to determine meson$-NN^{\ast }(1440)$ coupling
constants. Taken into account that the quark transition potential obtained
can be written at all distances in terms of baryonic degrees of freedom \cite
{16}, one way to obtain the coupling constants is to parametrize the
asymptotic central interactions as 
\begin{equation}
V_{NN\rightarrow NN^{\ast }(1440)}^{OPE}(R)=\frac{1}{3}\,\frac{g_{\pi NN}}{%
\sqrt{4\pi }}\,\frac{g_{\pi NN^{\ast }(1440)}}{\sqrt{4\pi }}\,\frac{m_{\pi }%
}{2M_{N}}\,\frac{m_{\pi }}{2(2M_{r})}\frac{\Lambda ^{2}}{\Lambda ^{2}-m_{\pi
}^{2}}[(\vec{\sigma}_{N}.\vec{\sigma}_{N})(\vec{\tau}_{N}.\vec{\tau}_{N})]\,%
\frac{e^{-m_{\pi }R}}{R}\,,
\end{equation}
and 
\begin{equation}
V_{NN\rightarrow NN^{\ast }(1440)}^{OSE}(R)=-\,\frac{g_{\sigma NN}}{\sqrt{%
4\pi }}\,\frac{g_{\sigma NN^{\ast }(1440)}}{\sqrt{4\pi }}\,\frac{\Lambda ^{2}%
}{\Lambda ^{2}-m_{\sigma }^{2}}\,\frac{e^{-m_{\sigma }R}}{R}\,.
\end{equation}
By comparing these baryonic potentials with the asymptotic behavior of the
ones previously calculated from the quark model, we can extract the $\pi
NN^{\ast }(1440)$ and $\sigma NN^{\ast }(1440)$ coupling constants in terms
of the elementary $\pi qq$ coupling constant and the one-baryon model
dependent structure.

The $[\Lambda ^{2}/({\Lambda ^{2}-m_{i}^{2}})]$ vertex factor in Eqs. (3)
and (4) comes from the vertex form factor chosen at momentum space at the
quark level, $[\Lambda ^{2}/({\Lambda ^{2}+\vec{q}^{\,\,2}})]^{1/2}$, where
chiral symmetry requires the same form for pion and sigma. Then it is clear
that the extraction from any model of the meson-baryon-baryon coupling
constants depends on this choice. We shall say they depend on the coupling
scheme.

To get $g_{\pi NN^{\ast }(1440)}/\sqrt{4\pi }$ we take our results for the $%
^{1}S_{0}$ one-pion exchange potential and we fit its asymptotic behavior
(in the range $R:5\rightarrow 9$ fm) to Eq. (3). We obtain 
\begin{equation}
\frac{g_{\pi NN}}{\sqrt{4\pi }}\frac{g_{\pi NN^{\ast }(1440)}}{\sqrt{4\pi }}%
\frac{\Lambda ^{2}}{\Lambda ^{2}-m_{\pi }^{2}}=\,-\,3.73\,,
\end{equation}

\noindent i.e. $g_{\pi NN^{\ast }(1440)}/\sqrt{4\pi }=-0.94$. As explained
above only the absolute value of this coupling constant is well defined. In
Ref. \cite{17} a different sign with respect to our coupling constant is
obtained what is a direct consequence of the different global sign chosen
for the $N^{\ast }(1440)$ wave function. The coupling scheme dependence can
be explicitly eliminated if we compare $g_{\pi NN^{\ast }(1440)}$ with $%
g_{\pi NN}$ extracted from the $NN\rightarrow NN$ potential within the same
quark model approximation. Thus we get 
\begin{equation}
\left| \frac{g_{\pi NN^{\ast }(1440)}}{g_{\pi NN}}\right| =0.25\,.
\label{eq17}
\end{equation}
This ratio is similar to that obtained in Ref. \cite{17} and a factor 1.5
smaller than the one obtained from the analysis of the partial decay width.
Nonetheless one can find in the literature values for $f_{\pi NN^{\ast
}(1440)}$ ranging between 0.27$-$0.47 coming from different experimental
analyses with uncertainties associated to the fitting of parameters \cite
{1,3,17}.

By proceeding in the same way for the OSE potential we obtain 
\begin{equation}
\left| \frac{g_{\sigma NN^{\ast }(1440)}}{g_{\sigma NN}}\right| =0.47\,.
\label{eq18}
\end{equation}
Our result agrees quite well with the only experimental available result,
obtained in Ref. \cite{18}, $0.48$. Furthermore, we can give a very
definitive prediction of the magnitude and sign of the ratio of the two
ratios,

\begin{equation}
\frac{g_{\pi NN^{\ast }(1440)}}{g_{\pi NN}}=0.53\;\frac{g_{\sigma NN^{\ast
}(1440)}}{g_{\sigma NN}}\,,
\end{equation}
which is an exportable prediction of our model.

\section{Acknowledgments}

The authors thank Dr. S. Hirenzaki for useful correspondence concerning the
calculation of Ref. \cite{13}. This work has been partially funded by
Ministerio de Ciencia y Tecnolog{\'{\i}}a under Contract No. BFM2001-3563,
by Junta de Castilla y Le\'{o}n under Contract No. SA-109/01, and by EC-RTN
(Network ESOP) under Contracts No. HPRN-CT-2000-00130 and HPRN-CT-2002-00311.

\end{document}